**Femtosecond profiling of shaped X-ray pulses**

M. C. Hoffmann[1*], I. Grguraš[2,3*], C. Behrens[4], C. Bostedt[1,5], J. Bozek[1,6], H. Bromberger[2,3], R. Coffee[1], J.T. Costello[7], L.F. DiMauro[8], Y. Ding[1], G. Doumy[5], W. Helml[1,10], M. Ilchen[1,11], R. Kienberger[9,10], S. Lee[12], A.R. Maier[3,13], T. Mazza[11], M. Meyer[11], M. Messerschmidt[1,14], S. Schorb[1], W. Schweinberger[9], K. Zhang[8], A.L. Cavalieri[2,3§]

[1] *SLAC National Accelerator Laboratory, 2575 Sand Hill Rd., Menlo Park, CA 94025, USA.*

[2] *Max-Planck Institute for the Structure and Dynamics of Matter, Luruper Chaussee 149, 22761 Hamburg, Germany.*

[3] *Center for Free-Electron Laser Science (CFEL), Luruper Chaussee 149, 22761 Hamburg, Germany*

[4] *Deutsches Elektronen-Synchrotron DESY, Notkestr. 85, 22607 Hamburg, Germany.*

[5] *Argonne National Laboratory, 9700 S Cass Ave, Lemont, Illinois 60439, USA*

[6] *Synchrotron SOLEIL, l'Orme des Merisiers, Saint-Aubin BP48, 91192 GIF-sur-YVETTE CEDEX, France*

[7] *School of Physical Sciences and National Center for Plasma Science and Technology (NCPST), Dublin City University, Glasnevin, Dublin 9, Ireland.*

[8] *Department of Physics, The Ohio State University, Columbus, Ohio 43210, USA*

[9] *Max-Planck-Institut für Quantenoptik, Hans-Kopfermann-Straße 1, 85748 Garching, Germany*

[10] *Physik-Deapartment E11, TU München, D-85748 Garching, Germany*

[11] *European XFEL GmbH, Albert-Einstein-Ring 19, 22761 Hamburg, Germany*

[12] *Korea Research Institute of Standards and Science (KRISS), Daejeon 305-600, Korea*

[13] *University of Hamburg, Institute of Experimental Physics, Luruper Chaussee 149, 22761 Hamburg, Germany.*

[14] *National Science Foundation BioXFEL Science and Technology Center, 700 Ellicott St., Buffalo, New York 14203, USA*

[*] *These authors contributed equally to this work*

[§] *corresponding author*

**Abstract**

Arbitrary manipulation of the temporal and spectral properties of X-ray pulses at free-electron lasers (FELs) would revolutionize many experimental applications. At the Linac Coherent Light Source at Stanford National Accelerator Laboratory, the momentum phase-space of the FEL driving electron bunch can be tuned to emit a pair of X-ray pulses with independently variable photon energy and femtosecond delay. However, while accelerator parameters can easily be adjusted to tune the electron bunch phase-space, the final impact of these actuators on the X-ray pulse cannot be predicted with sufficient precision. Furthermore, shot-to-shot instabilities that distort the pulse shape unpredictably cannot be fully suppressed. Therefore, the ability to directly characterize the X-rays is essential to ensure precise and consistent control. In this work, we have generated X-ray pulse pairs and characterized them on a single-shot basis with femtosecond resolution through time-resolved photoelectron streaking spectroscopy. This achievement completes an important step toward future X-ray pulse shaping techniques.

Free-electron lasers (FELs) operating from the extreme ultraviolet (XUV) to the hard X-ray spectral regime emit femtosecond pulses that are nearly ten orders of magnitude brighter than pulses generated by any other femtosecond X-ray source.[1,2,3,4] Such intense pulses have enabled new classes of experiments across a broad range of disciplines[5] in the natural sciences including structural biology,[6,7,8] femtochemistry,[9,10] solid-state physics,[11,12,13] and high energy density science.[14,15] To rapidly build on proof-of-principle experiments and initial demonstrations of new techniques, even greater control over the X-ray pulse properties is desirable.

Manipulation of the driving electron beam presents a clear route to gain control over the FEL emission. For example, at X-ray FELs based on self-amplified-spontaneous-emission (SASE), the duration of the FEL pulse is dependent on, and in most cases fundamentally limited by, the length of the driving electron bunch.[16] Therefore, shorter X-ray pulses can be realized by stronger electron bunch compression. With this approach, pulses of less than 10 femtoseconds have been generated.[17]

However, as the peak current in the FEL is limited by Coulomb repulsion and other collective effects, stronger compression requires a reduction in the total charge of the FEL driving electron bunch.

Alternatively, a simpler approach for generating even shorter X-ray pulses involves manipulating the electron beam emittance, which is a figure of merit of the electron beam quality. If the emittance is "spoiled" in a controlled fashion in selected regions of the electron bunch, only the "unspoiled" slices of the bunch will have high enough quality to support the SASE process.[18] It was originally predicted and later proven experimentally that sub-5 femtosecond X-ray pulses could be generated using this electron bunch slicing method.[19,20] Moreover, it is expected that attosecond X-ray pulses can be generated if this technique is applied in the hard X-ray regime.

Beyond slicing for control over the X-ray pulse duration, it was proposed that the spoiler scheme could be extended to produce pairs of X-ray pulses from a single driving electron bunch. This is particularly attractive as effective methods for producing X-ray pulse pairs based on traditional interferometer geometries using beam splitting and recombination are difficult to implement. In fact, multiple approaches have been devised to circumvent this experimental difficulty.[21,22] Here, with the "spoiler", it is straight forward to produce X-ray pulse pairs by preserving the emittance in two distinct regions of the electron bunch. By adjusting the extent of these regions, the durations of the individual X-ray pulses can be tuned. Additionally, varying the separation between the two unspoiled regions permits precise setting of the delay between the emitted X-ray pulses. Furthermore, even the photon energy of the individual X-ray pulses can be controlled by independently tuning the final electron beam energy of the two unspoiled electron bunch slices.

X-ray pulses with variable delay enable X-ray pump, X-ray probe experiments,[23,24,25] allowing deeply bound core states or highly excited conduction band states to be selected and probed in a fully controlled manner. Independent control over the X-ray photon energies combined with precise time-delays between the two X-ray pulses will allow for many novel approaches to be applied across the full range of disciplines in ultrafast science.

Although almost any parameter of the multi-dimensional electron bunch phase-space can be varied, in practice, without precise measurement, it is not exactly clear how the composition of the delivered electron bunch is affected by such parameters, and thus the highly nonlinear process of FEL emission cannot be predicted with certainty. Furthermore, even if the electron beam was to be perfectly characterized at the entrance to the X-ray undulators (a diagnostic challenge that has not yet been met) these are typically destructive measurements[26] that are of limited value when considering drifts and shot-to-shot fluctuations. Therefore, precise temporal diagnostics for the FEL photon pulse are crucial for the success and application of this "spoiler" technique and for X-ray pulse shaping techniques in general.

In this work, we demonstrate all key elements required for advanced X-ray pulse shaping through electron beam manipulation by first generating X-ray pulse pairs and then directly characterizing the X-rays to verify, control and tune the delivered pulse profile.

**Electron Bunch Manipulation**

In the "spoiler" scheme, the beam emittance is accessed in a magnetic chicane, which is a dispersive section of the accelerator, and is analogous to a prism compressor used in ultrafast optics. In the chicane, as depicted in Fig. 1, high-energy electrons follow shorter trajectories than low energy electrons, which results in longitudinal compression of bunches that are initially linearly chirped in energy. Energy chirp is introduced or eliminated by appropriate phasing of the accelerating fields. In the middle of the chicane, the longitudinal, or temporal coordinate of the electron bunch phase-space is transformed into a transverse spatial coordinate $x$. At the Linac Coherent Light Source (LCLS), part of the SLAC National Accelerator Laboratory, femtosecond X-ray pulse pairs can be generated by inserting a thin metallic foil with two slots at this position. The regions of the bunch that pass through the openings in the "double-V-slotted" foil propagate through the chicane with preserved emittance and are expected to lase normally in the FEL undulator. In contrast, Coulomb scattering of the electrons in the regions of the bunch that collide with the opaque parts of the foil causes an increase in the transverse emittance, leading to strong suppression of FEL gain. As a result, upon recompression and homogenization of the beam energy (to eliminate the residual energy chirp) the two unspoiled slices of the electron bunch generate a pair of collinear X-ray pulses with finite delay. In principle, for a double-slotted foil with slot separation $\Delta x$, the delay between the two unspoiled slices upon recompression can be estimated from[27]:

$$\Delta t = \frac{\Delta x}{\eta h C c} \quad , (1)$$

where $\eta$ is the momentum dispersion in the middle of the chicane, $h$ is the degree of linear energy chirp introduced in the accelerator section prior to the magnetic chicane, $C$ is the bunch compression factor, and $c$ the speed of light in vacuum. Crucial beam parameters including bunch compression factor and initial chirp are difficult to calculate and are challenging to measure experimentally. Furthermore, it is not always accurate or desirable to assume that the bunch is initially linearly chirped. Even more important, due to shot-to-shot fluctuations it is not always clear that the "unspoiled" parts of the bunch that pass through the slots in the foil will ultimately lead to amplified FEL emission, which is a highly nonlinear process.

Indirect observation of the X-ray pulse structure has been achieved by analysis of the electron bunch after it has propagated through the FEL undulator.[17] Here, the energy loss of the electrons participating in the FEL emission process can be determined using a transverse deflection cavity (XTCAV). This information is then used to deduce the X-ray photon pulse profile. An advantage of this measurement is that it can be made without affecting the X-ray emission. However, for very short X-ray pulses, especially sub-femtosecond X-ray pulses that may be generated in the future, the technique suffers from mismatch in the effective velocity of the electron beam in the undulator structure and the speed of light, which fundamentally limits the measurement resolution. Furthermore, this beam-based diagnostic provides an indication of X-ray pulse structure only at the point of emission and not at the locus of the actual experiment. Therefore, it not applicable in future X-ray pulse shaping schemes that will utilize dispersive X-ray optics nor can it be used to account for unwanted pulse distortion due to X-ray transport optics, for example. These limitations emphasize the importance of direct photon-based diagnostics like THz streaking used in this work.[28,29]

**Time-resolved photoelectron streaking spectroscopy**

Single-cycle THz pulses were used to directly characterize the X-ray photon pulses produced at LCLS. An overview of the THz streaking apparatus used for these measurements is shown schematically in Fig. 2. The X-ray pulses are focused onto a neon gas target and temporally and spatially overlapped with an intense linearly polarized THz pulse. The FEL pulse ionizes the gas target, producing a burst of photoelectrons with a temporal profile that replicates the profile of the incident photon pulse. The final kinetic energy of the electrons that compose the photoelectron burst is subsequently increased or decreased depending on their exact time of release into the THz field. The final kinetic energy, $E_\mathrm{f}$, for an electron released at $t_0$ and observed parallel to the THz electric field, in atomic units, is given by[30]:

$$E_\mathrm{f}(t_0) = E_i - p_i A(t_0) + \frac{A^2(t_0)}{2} \, . \quad (2)$$

Here $E_i$ and $p_i$ are the initial kinetic energy and momentum, respectively, of the electron, and $A(t_0)$ is the THz vector potential at the instant of ionization. For a burst of photoelectrons emitted over a finite period of time, the photoelectron spectrum is broadened depending on the duration of emission and gradient of the streaking field. When the duration of the ionizing X-ray pulse is shorter than the half-cycle of the THz streaking field, the FEL pulse structure can be reconstructed from the measured photoelectron spectrum provided that the instantaneous THz vector potential is also known.

**Characterization of X-ray pulse pairs**

X-rays delivered at LCLS were characterized at the Atomic, Molecular and Optical (AMO) end station. Two sets of experiments were performed using different methods of THz generation to optimize the dynamic range and measurement resolution. In the first set of experiments, single-cycle THz pulses were generated by rectification of near-infrared laser pulses in lithium niobate (LN) by the tilted pulse-front method.[31] These pulses had a streaking field half-cycle of ~ 660 fs, which gives an upper bound on the dynamic range of the measurement. While 660 fs is long enough to accommodate both the hundred-femtosecond timing jitter at LCLS [32,33] and extended X-ray emission, the time resolution that can be achieved is limited to approximately 50 fs FWHM.

Therefore, in the second set of experiments, designed to measure shorter X-ray pulses and pulse structures below 50 fs, THz pulses were generated by rectification of infrared (IR) laser pulses in the organic crystal DSTMS.[34] The resulting THz pulse has a frequency spectrum centred at ~ 3

THz, compared to a spectrum centred at approximately 1 THz for pulses generated in LN. With higher frequency components, the THz streaking field is naturally steeper, providing higher time-resolution. However, the dynamic range of the measurement in this case is only ~180 fs, which no longer accommodates the timing jitter at LCLS. To overcome this limitation, additional independent timing information is provided with ~10 fs rms accuracy using the technique of spectral encoding[35,36] as described in the methods section.

In the experiments, as shown in Fig. 2, THz pulses generated by either LN or DSTMS are overlapped with X-ray pulses with a photon energy of ~1.0 keV and focused in a neon gas target where the 1s core level is ionized. To establish temporal overlap, a series of single-shot photoelectron spectra is recorded as the set relative delay between standard, unshaped X-ray pulses and LN-THz pulses is scanned. The spectrogram shown in Fig. 3a is generated by averaging a series of sequential single-shot acquisitions. Here, the vector potential of the LN-THz pulse is clearly discernible, allowing the peak field strength to be accurately determined, which is crucial for calibration (see Methods). Next, the X-ray pulse was shaped by manipulating the electron bunch emittance in the magnetic chicane with the double-slotted spoiler. As shown schematically in Fig. 1, the spoiler has a V-shape so that the delay between X-ray pulse pairs can be tuned by varying the insertion depth of the foil. To verify control over the emitted pulse shape and delay between pulses, X-rays were observed as the foil was scanned.

A characteristic measurement for one foil position is shown in Fig. 3, the streaked photoelectron spectrum is shown in Fig. 3b, and the retrieved temporal profile is shown in Fig. 3c. Here the observed delay between the X-ray pulses is ~ 145 fs. Upon scanning the insertion depth, the maximum delay between the X-ray pulses was observed to be (215 ± 21) fs, while the minimum delay observed with this accelerator configuration was (65 ± 10) fs. As expected, the delay between the individual X-ray pulses in the double-pulse depends linearly on the insertion depth. The errors in the measured delays here and throughout the paper are determined from the numerical distribution of observed single-shot delays.

A crucial point, however, is that not all measurements revealed the double pulse structure in the streaked photoelectron spectrum. This indicates that the spoiler method is not guaranteed, highlighting the need for single-shot temporal diagnostics. The fraction of observed double-pulses was highest for the intermediate separation, while the double-pulse structure was less likely to be observed at the narrowest and furthest slot-separation. At the largest separation, when the double-pulse structure is not observed, it is believed that only one half of the X-ray pulse pair was produced. The production of only one half of the pulse pair may be due to inhomogeneity of the driving electron bunch in the far reaches of the leading and trailing edges of the electron bunch, which are the parts of the bunch that are unspoiled for this foil insertion depth in the magnetic chicane. In contrast, at the narrowest separation, where the delay is near the streaking measurement resolution, failure to observe the X-ray pulse pair is due to experimental limitation rather than a true absence of structure in the X-ray emission. This conclusion is supported by the second set of THz streaking measurements that were made with higher time-resolution.

Using DSTMS-generated THz pulses, with a rise time of ~ 180 fs, time resolution of ~20 fs FWHM was achieved. This value corresponds to the minimum separation between distinguishable peaks in an observed photoelectron spectrum and not the accuracy with which we can determine the separation between observed peaks. With this higher temporal resolution, LCLS was reconfigured to emit X-ray pulse pairs with smaller delay, which requires a higher compression factor in the compressor chicane. Under these conditions a spectrogram recorded at an intermediate slot-separation is shown in Fig. 4. By incorporating the independent arrival time information provided by the spectral timing tool, the averaged spectrogram can be constructed with an effective timing jitter of only 10 fs rms. As a result, the signal splits into two distinct curves around the zero-crossing of the THz streaking field, indicating the persistence of X-ray pulse pairs. Since the streaking effect is imprinted on both the positive and negative slope of the vector potential it should be possible to recover additional parameters such as the average chirp or spectral phase of individual pulses with our THz streaking technique.[30] Analysing the separation between the two THz vector potential curves in the averaged spectrogram at the field's zero-crossing gives a delay between the X-ray pulses of 56 fs. Complementary analysis of a single-shot measurement is plotted in Fig. 4b and Fig. 4c, which yields a delay between the X-ray pulses of ~ 54 fs, in agreement with the averaged measurement. In this set of experiments, the double-slotted spoiler and accelerator configuration were tuned to generate X-ray pulse pairs with a maximum measured time delay of (66 ± 4) fs and a minimum measured delay of (38 ± 6) fs.

It should be noted that the minimum delay observed with the DSTMS-generated THz was not limited by the measurement resolution. This is evidenced by the fact that the modulation depth between the two pulses at the minimum observed delay of 38 fs is greater than 50% of the peak maximum. Nor was the minimum observed delay limited by the accelerator. Rather, a configuration of the machine that would have resulted in a narrower separation was not explored in this work. Nevertheless, it can be expected that X-ray pulse pairs with shorter delay can be generated and that with the current THz diagnostic, these pulse pairs can be distinguished to a minimum separation of ~20 fs. As in the case for LN-based THz diagnostic, this limit is determined using the product of the field-free photoelectron bandwidth and the THz streaking strength.

The results of all pulse shaping and characterization measurements are summarized in Fig. 5. As previously mentioned, the delay between the X-ray pulses varies linearly with the separation between the slots in the spoiler. For comparison, the delay between the pulses can also be calculated. This requires simulating the linear accelerator configuration to obtain values for the momentum dispersion, chirp and compression factor, which are then substituted in Eq. 1. The calculation is overlaid on the experimental data points in Fig. 5. There is excellent agreement between the measured and calculated values, indicating that the machine parameters are relatively well known for this simple double-slotted spoiler case.

**Conclusion**

THz streaking spectroscopy was used to verify paired X-ray pulse production at LCLS using the double-slotted spoiler. In the future, this diagnostic tool could be used to optimize and diagnose the pair pulse production facilitating X-ray pump/ X-ray probe experiments. While this is a basic demonstration of the concept of X-ray pulse shaping by electron beam manipulation, these advances may lead to more sophisticated, spectrally and temporally tailored pulse shapes for targeted experimental applications. Electron bunch manipulation, in combination with other new technologies, may even lead to intense isolated attosecond X-ray pulse generation with photon energy that can be freely tuned into the hard X-ray regime. Further into the future, reliable control over the electron beam momentum phase space would allow for emission of chirped X-ray pulses that could then be recompressed with dispersive X-ray optics. In this way, the methods of chirped pulse amplification that have been crucial to the development of high-power ultrafast laser science and technology could be extended to the hard X-ray regime.[37]

The key to these and other future developments in machine and experimental applications will be the ability to temporally characterize the emitted FEL X-ray pulse. Here, using THz waveforms for streaking spectroscopy, a measurement resolution of ~20 fs FWHM has been achieved. By using stronger THz fields, or shorter streaking wavelengths in the mid-IR and IR, the resolution can be improved to accommodate sub-femtosecond X-ray pulses or pulses with substructure on the attosecond time scale.

**Acknowledgements**

Portions of this research were carried out at the Linac Coherent Light Source (LCLS) at the SLAC National Accelerator Laboratory. The LCLS is an Office of Science User Facility operated for the US Department of Energy Office of Science by Stanford University. DCU contribution was made possible under Science Foundation Ireland IvP Grant No. 12/IA/1742 and the EU EMJD 'EXTATIC' under framework partnership agreement FPA-2012-0033. M.I. acknowledges funding from the Volkswagen foundation within the Peter Paul Ewald-Fellowship. WH, RK & WS acknowledge funding from the Bavaria California Technology Center (BaCaTeC). MMes acknowledges support by NSF award 1231306. KZ and GD acknowledge support from US NSF grant PHY-1004778 and US DOE grant DE-FG02-04ER15614. SL acknowledge support from NRF-2014M3C1A8048818, NRF-2014M1A7A1A01030128, NRF-2016K1A3A7A09005386

The authors would like to thank Steve Edstrom for expert laser support and Nick Hartmann for help with the spectral encoding time-tool.

**Author Contributions:**

A.L.C. and M.C.H. conceived the experiment. C.B., Ch. B., J.B., H.B. A.L.C., R.C., J.T.C., L.F.D., Y.D., G.D., I.G., W.H., M.C.H, M.I., R.K., S.L., A.R.M., T.M., M.M., M.Mes, S.S., W.S. and K.Z. performed the experiment. Y.D. implemented the double-slotted spoiler. R.C. devised spectral encoding time-tool. I.G., H.B., A.L.C., S.L. and T.M. performed the data analysis. I.G., C.B., A.L.C., J.T.C., L.F.D., M.C.H and M.M. contributed significantly in the preparation of the manuscript.

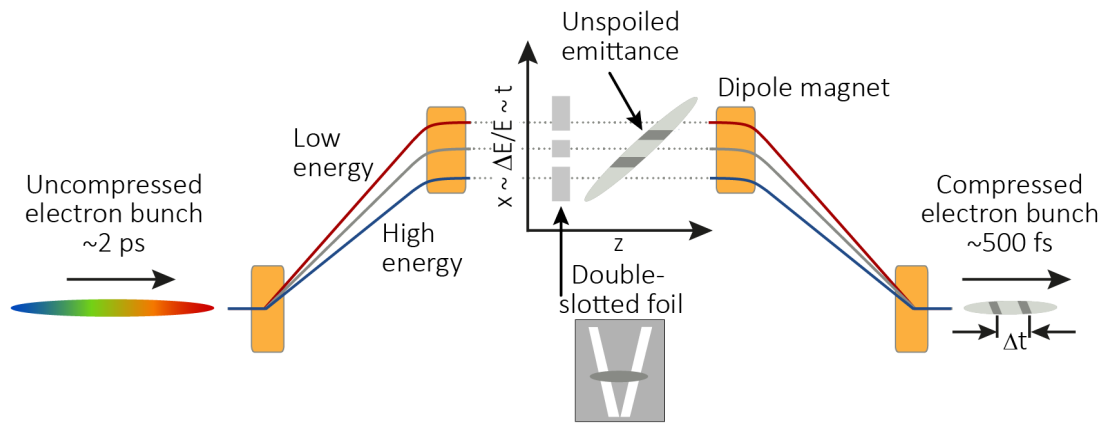

**Figure 1| Spoiled emittance for X-ray pulse shaping.** Chirped electron bunches are compressed in a magnetic chicane, which is analogous to a prism compressor in ultrafast optics. To generate X-ray pulse pairs, a "V-shaped" slotted aluminium foil is inserted in the centre of the magnetic bunch compressor chicane, where the electrons are maximally dispersed in the transverse dimension. The electron beam emittance is preserved for those parts of the bunch that pass through the slots, while the emittance is spoiled for the rest. Upon compression at the exit of the chicane the two unspoiled parts are separated in time by an amount that depends on the lateral separation between the slots in the foil, resulting in the generation of two X-ray pulses in the FEL undulator. Residual energy chirp is eliminated by higher phasing of the fields in the final accelerating modules, as well as higher order effects in the accelerator structure.

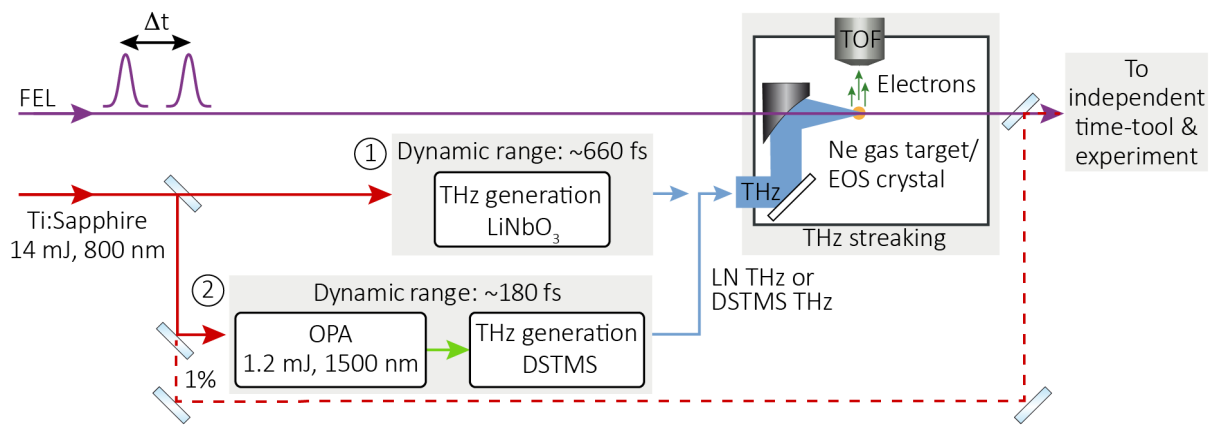

**Figure 2| Direct measurement of X-ray pulse pairs.** X-ray pulses are temporally characterized on a single-shot basis by THz streaking. Single-cycle THz pulses were generated either by the tilted pulse-front method in lithium niobate (LN) or by collinear optical rectification of infrared pulses in the organic crystal DSTMS. The two different THz pulses were used to optimize the trade-off between the dynamic range and temporal resolution in the measurement. The X-ray FEL and THz pulses are focused and spatially and temporally overlapped in a neon gas target. The kinetic energy of the photoelectrons ionized by the X-ray FEL pulse and streaked with the THz field is measured with a time-of-flight (TOF) spectrometer in plane with the X-ray and THz polarization. For calibration of the retrieved X-ray pulse profile, the Ne gas target can be replaced with a nonlinear crystal for independent characterization of the THz electric field by electro-optic sampling (EOS) (while X-rays are not being measured). If required due to excessive jitter, a spectral encoding time-tool can be used to more precisely determine the temporal overlap between the X-ray and THz pulses.

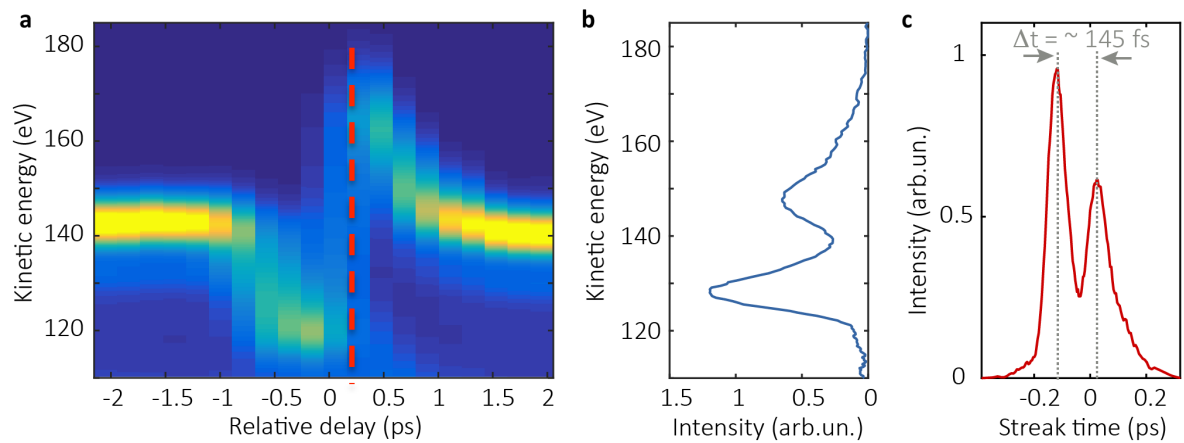

**Figure 3| Temporal characterization with LN-THz pulses.** The spectrogram shown in panel **a** contains streaked neon 1s photoelectron spectra that are collected on a single-shot basis and averaged according to the set relative delay between the X-ray FEL and THz pulses. The X-ray photon energy was 1.01 keV. Clear observation of the THz vector potential is used for temporal overlap and the maximally shifted streaked spectra at the peaks of the vector potential are used to calibrate the streaking diagnostic. A characteristic single-shot spectrum taken around 0.2 ps delay (dashed line) is shown in panel **b**. The double X-ray pulse structure is clearly observed. The corresponding X-ray temporal pulse profile is retrieved and shown in panel **c**. In this case, the two X-ray pulses are separated by ~ 145 fs.

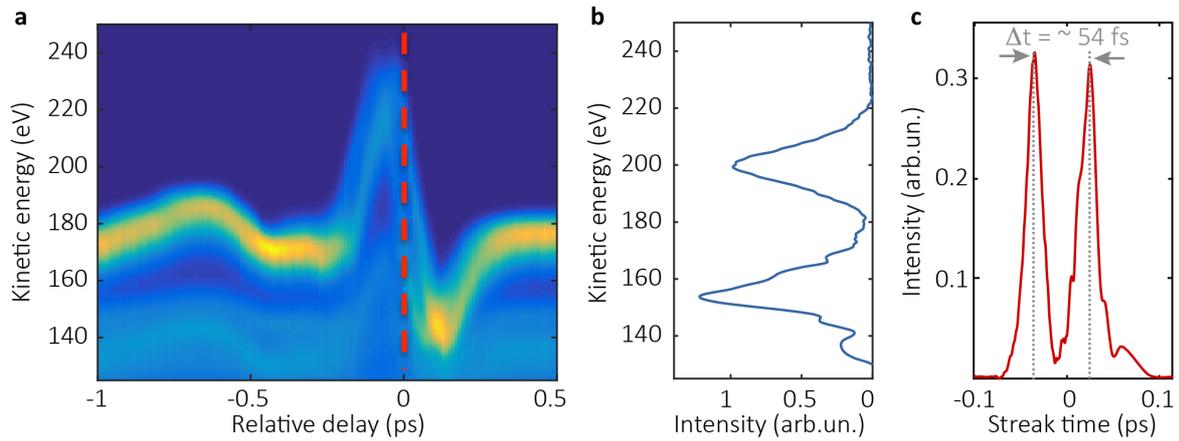

**Figure 4| High resolution temporal characterization with DSTMS-THz pulses.** The spectrogram shown in panel **a** contains single-shot spectra averaged according to the set relative delay between the X-ray FEL and THz pulses, plus a correction provided by the spectral encoding time-tool (see Methods). The X-ray photon energy was 1.04 keV. Due to the X-ray pulse structure imprinted by the double-slotted spoiler, the streaked photoelectron spectrum splits into two distinct peaks near time-zero. A characteristic single-shot spectrum at zero delay is shown in panel **b**. The corresponding X-ray temporal pulse profile is retrieved and shown in panel **c**. In this case, the two X-ray pulses are separated by ~ 54 fs.

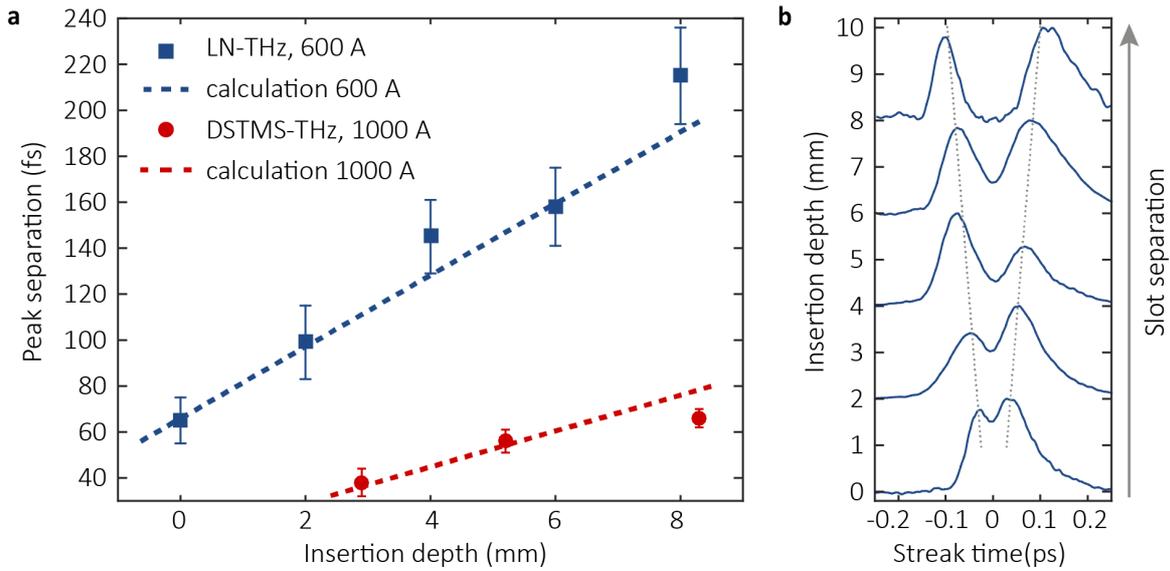

**Figure 5| Control over the X-ray pulse structure.** The double-slotted spoiler was scanned for two different compression factors in the magnetic chicane (600 A and 1000 A). LN-THz is used to measure X-ray pulses generated with the lower compression factor, and DSTMS-THz (with higher temporal resolution) to measure pulses with the higher compression factor. In panel **a**, as expected, the delay between the X-ray pulse pairs varies linearly with the lateral distance between the slots in the emittance-spoiling foil. In practice, the separation is controlled with the insertion depth of the foil in the centre of the chicane. For these simple X-ray pulse shapes, the calculated delay based on simulation of the linear accelerator is in excellent agreement with the observed delay. Panel **b** shows characteristic single-shot measurements obtained using the lower compression setting at 5 different insertion depths. The dotted lines indicate the expected linear dependence.